# Accessibility in 360º video players

**Chris. J. Hughes**

School of Computer Science, University of Salford, Manchester, UK, c.j.hughes@salford.ac.uk, orcid.org/0000-0002-4468-6660

**Mario Montagud**

i2CAT Foundation, Barcelona, Spain; University of Valencia, Spain, mario.montagud@i2cat.net

## ABSTRACT

Any media experience must be fully inclusive and accessible to all users regardless of their ability. With the current trend towards immersive experiences, such as Virtual Reality (VR) and 360º video, it becomes key that these environments are adapted to be fully accessible. However, until recently the focus has been mostly on adapting the existing techniques to fit immersive displays, rather than considering new approaches for accessibility designed specifically for these increasingly relevant media experiences.

This paper surveys a wide range of 360º video players and examines the features they include for dealing with accessibility, such as Subtitles, Audio Description, Sign Language, User Interfaces, and other interaction features, like voice control and support for multi-screen scenarios. These features have been chosen based on guidelines from standardization contributions, like in the World Wide Web Consortium (W3C) and the International Communication Union (ITU), and from research contributions for making 360º video consumption experiences accessible. The in-depth analysis has been part of a research effort towards the development of a fully inclusive and accessible 360º video player. The paper concludes by discussing how the newly developed player has gone above and beyond the existing solutions and guidelines, by providing accessibility features that meet the expectations for a widely used immersive medium, like 360º video.

## Funding

This work has been partially funded by the European Union's Horizon 2020 program, under agreement nº 761974 (ImAc project). Work by Mario Montagud has been additionally funded by the Spanish Ministry of Science, Innovation and Universities with a Juan de la Cierva – Incorporación grant (with reference IJCI-2017-34611).

## Code availability

The ImAc player is open source, and available from https://github.com/ua-i2cat/ImAc under a GPL-3.0 license.

## NOTE:

This article is currently under review process in Multimedia Tools and Applications (MTAP) journal. The submission to Arxiv has been allowed by the MTAP Editors handling the review process.

# 1  Introduction

The recent explosion of immersive media technologies, like Virtual Reality (VR) and 360º video, opens the door to new fascinating opportunities and revenue models, not only in the entertainment sector, but also in other key sectors of society, like education and culture [1]. In this context, 360º videos have become a simple and cheap, yet effective and hyper-realistic, medium to provide VR experiences. Due to their potential, the scientific community and industry have devoted significant efforts in the last years to providing improved solutions in terms of many relevant aspects, like capturing and consumption hardware, compression strategies, representation formats, and media players. Likewise, the demand for production and consumption of 360º videos has significantly increased, and major platforms, like Youtube, Facebook and New York Times, currently provide 360º videos in their service offerings [1]. This has also led to the development of a wide range of 360º players for a variety of platforms and consumption devices, like desktop computers, smartphones and Head Mounted Displays (HMDs) [2, 3].

As for every service, 360º media consumption experiences need to be accessible. Typically, accessibility has been considered in the media sector as an afterthought, and mainly for mainstreamed services. The community is fully aware of the relevance of accessibility and of the existing regulation frameworks to guarantee accessible services. However, it seems that accessibility for immersive media services is still in its infancy. Likewise, the lack of standardized solutions and guidelines has catered for the development of own non-unified solutions, meeting specific requirements.

This situation has served as a motivation to conduct the research study presented in this work, which is *an in-deep exploration and categorization of how and to what extent accessibility services are integrated in the key existing 360º players in the media landscape*. This main contribution of the paper is an initial, but necessary step, towards identifying what are the advances and limitations in this field. Based on the insights from the conducted survey and analysis, a second contribution of the paper is the *description of the iterative process towards a user-centric design and development of a new 360º player* [4, 5] that aims at filling the existing gaps, providing tested accessibility features based on users' needs and preferences.

## 1.1 Focus and Target Users

Everyone has the right to access and comprehend the information provided by media services in general, and by immersive 360º experiences in particular. Therefore, accessibility becomes a priority, not only to adhere to current worldwide regulations, but also to reach a wider audience and contribute to equal opportunities and global e-inclusion. Immersive experiences need to be fully inclusive across different languages, addressing not only the needs of consumers with hearing and vision impairments, but also of those with cognitive and/or learning difficulties, low literacy, newcomers, and the aged.

Among the requirements to enable universal access to immersive 360º content, this paper focuses on solutions to achieve an efficient integration of access services (like subtitling, audio description and sign language) with 360º content, and on appropriate User Interfaces (UIs) to mostly enable an effective interaction and usage of these services.

The studies in [1 6] provide statistics about the percentages of the worldwide population with some form of audio-visual impairments, and about the increasing ageing process [7], which is also strongly related to accessibility needs. In addition, these studies reflect on the societal impact of media accessibility, and review the existing regulatory framework to ensure a full democratic participation in the current society where technologies and media play a key role.

Therefore, the current situation with regard to accessibility needs, in combination with the associated regulatory framework, are raising awareness and putting pressure on content producers and providers for fulfilling the missing requirements. On the one hand, the presented study and categorization serve to motivate the need for further advancing on the accessibility field for immersive media. On the other hand, the contributions of the paper are meant to become a valuable resource for the interested audience in this field, including: 1) users with accessibility needs, in order to select the player that best fits their needs; 2) the development community and service providers, in order to improve their solutions; and 3) the standardization bodies and research community, in order to have an overall view of what is solved and what is missing, and/or determine to what extent the existing guidelines are met.

## 1.2 Accessibility Guidelines

The World Wide Web Consortium (W3C) provides detailed guidelines for producing content for Web environments, such as the *Web Content Accessibility Guidelines* (WCAG) [8], and the presented study assesses the level of conformity of existing 360º video players with these guidelines. In particular, the WCAG defines four key services that are required in order for a 360º video to be regarded as accessible:

1. A transcript, which is a written version of the spoken audio, and is required in order to provide the most basic level of accessibility.
2. Subtitles (ST), which are essentially the transcript, broken into small sections (usually ~30 characters, 2 lines), synchronized with the video and/or audio. Although not an essential requirement, ST are desirable for people who are deaf, hard of hearing or non-native speakers. It allows for the text to be read whilst also viewing facial expressions, body language and actions which are used to indicate the context and intent of the specific characters who are speaking. Usually, different colours are used to identify each speaker, and special characters are often used to identify sounds or actions. ST are particularly important when consuming immersive content, as reading a separate transcript document would break the immersion on a desktop computer, and even be impossible when wearing a HMD.
3. Audio Description (AD), which is an additional audio track played over the top of the main audio, and typically produced by a skilled audio describer to fit a description into the gaps in the main dialogue/action. The AD must provide a description for any relevant visuals which have not yet been discussed in the dialogue. The AD must describe what is being visually represented in order to address the needs of those who are blind or partially sighted. Technically, AD is commonly mixed with the main audio track/stream, which is not ideal for personalisation (a key aspect for accessibility).
4. Sign Language (SL), which uses the movement of the hands, facial expression and body language to convey meaning. Sign languages are full-fledged natural languages with their own grammar and lexicon and, as such, they are not universal and mutually intelligible with each other. Often, a signer is overlaid onto the video (Picture-in-Picture) in order to translate the dialog into a sign language. SL is not directly required for accessibility, although translation and interpretation is recommended for people who are deaf, and especially for whom sign language is their primary language. Technically, SL is commonly burned with the main video track/stream, which is not ideal for personalisation (a key aspect for accessibility).

These WCAG guidelines have become a widespread and useful measure of accessibility in a web context. In the broadcast sector, other more relaxed guidelines have been provided, e.g by International Telecommunication Union (ITU) [9]. In addition, an overview of contributions and initiatives towards standardizing accessibility guidelines can be found in [10].

The comparison aspects considered in this study are then based on the accessibility guidelines previously identified, which rely on the support for the traditional access services: ST, AD and SL. In addition, two further key aspects for accessibility are examined:
- support for Accessible UIs, and voice control interfaces, which become key for an effective usage of these services.
- support for multi-screen scenarios, which allow for richer personalisation and overcoming distance barriers to the main screen [11].

## 1.3 Research Methodology

This research study aims at shedding some light on the existence, lack and/or limitations with regard to accessibility solutions for 360º media content consumption. After defining the target users (Section 1.1) and selecting the accessibility guidelines and aspects to take into account (Section 1.2), a wide sample of existing 360º players have been surveyed, analyzed and categorized. The goal of the study is not to list all existing 360º players, but at least to compile and analyze a wide sample of them, including the key ones. The selection has been made by thoroughly searching for the existing 360º players, by analyzing the state of the art and existing online 360º video services, and by asking experts in the field, like content providers, broadcasters and researchers.

Then, a set of user-centric activities have been conducted to both gather accurate requirements and validate them. These activities will be briefly introduced in Section 4.

## 2   Overview of 360º Video Players

In this section, the key web-based and executable 360º video players are reviewed, and their approach to accessibility is discussed. Apart from the aspects identified in Section 1.2, the selection of the players have been also based on their support both desktop mode (in a web browser) and VR modes (using a HMD). The section starts by reviewing web-based players in Section 2.1, which are supposed to provide cross-platform, cross-device, and even cross-browser support, and then continues by reviewing executable players in Section 2.2, by detailing the supported platforms. A qualitative study and discussion about the pros and cons of executable vs web-based applications can be found in [12].

### 2.1 Web based 360º Players

**JWPlayer**[1] provides an Application Programming Interface (API) specifically targeted at developers and designers who are building their own apps and websites for delivering video. The API is very extensive and provides either a commercial version integrated with a Content Delivery Network (CDN) or a free to use version for self-hosting. In particular, the API enables the setup of a 360º player, and its embedding in a web context. The default interface of JWPlayer is shown (see Figure 1, left).. The basic UI is very standard for a video player, although a series of keyboard shortcuts are also provided.

As an API it is very limited in what it provides out of the box for accessibility. However, it does provide an extensible framework, and it is also possible to adapt the UI using JavaScript.

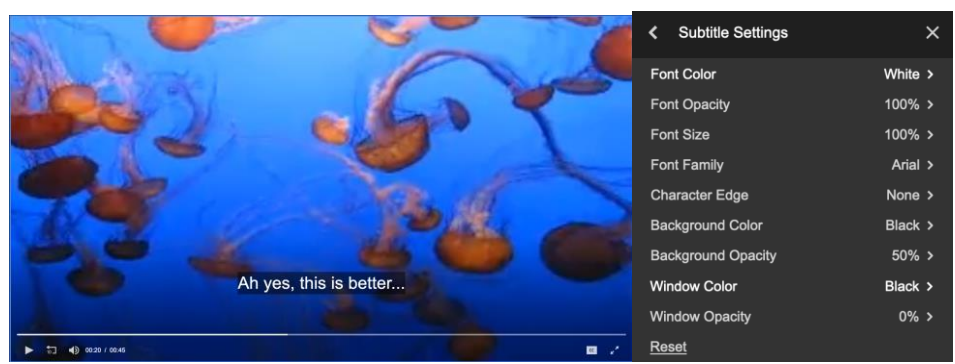

**Figure 1: JWPlayer: default interface (left), Subtitle Options (right)**

In terms of accessibility, JWPlayer supports three subtitle formats: Web Video Text Tracks (WebVTT) [13], SubRip Subtitle (SRT) [14] and Timed Text Markup Language (TTML) [15]. It also supports captions which are embedded in HTTP streams, like Dynamic Adaptive Streaming over HTTP (DASH) and HTTP Live Streaming (HLS). It provides the user with basic customisation controls for the rendering of the subtitle, but no control over position (see Figure 1, right).

Likewise, JWPlayer can support multiple audio tracks. Although there is technically no specific AD support, the support of multiple audio tracks can be used as a solution for delivering AD, by having an additional audio track which combines the main audio with AD. There is currently no support for SL, as it is generally accepted that the signer would be burned into the video, thus providing no option for personalisation.

---

[1] www.jwplayer.com - All links provided in the article have been accessed for the last time in May 2020

In addition, it is worth mentioning that there are open source projects for adding voice control support to JWPlayer, such as the integration of Siri on iOS devices [16].

**Omnivirt**[2] was primarily developed for commercially displaying advertising, although the company developing it has branched into 360º video. As a result of developing advertising banners which contain 360º video, a stand alone 360º video player has been developed. Omnivirt is free to use with some limitations (10k monthly views, 2GB files), after which you are required to upgrade to a premium version.

The UI of Omnivirt is very limited (see Figure 2) . However, it provides an interesting feature for 360º video consumption, which is a radar display indicating the current viewing position relative to the centre of the omnidirectional scene. This feature allows the user to recentre the view by clicking on the radar icon. The **Omnivirt** player does not provide support for any of the access services considered in the study.

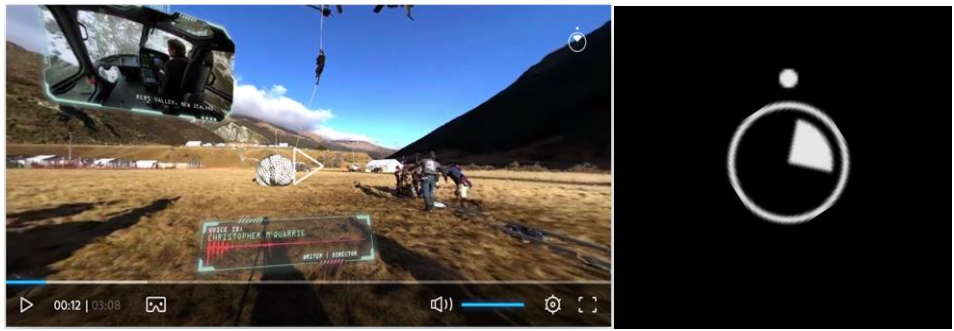

**Figure 2: The Omnivirt UI (left), with a cropped view of the radar display showing the users position within the scene (right).**

**YouTube**[3] is one of the largest players in the video hosting and delivery sector, which has also moved into the 360º video space. YouTube allows users to upload their own content which is then shared through its network. Although it is free to use, revenue is made by placing adverts on the videos.

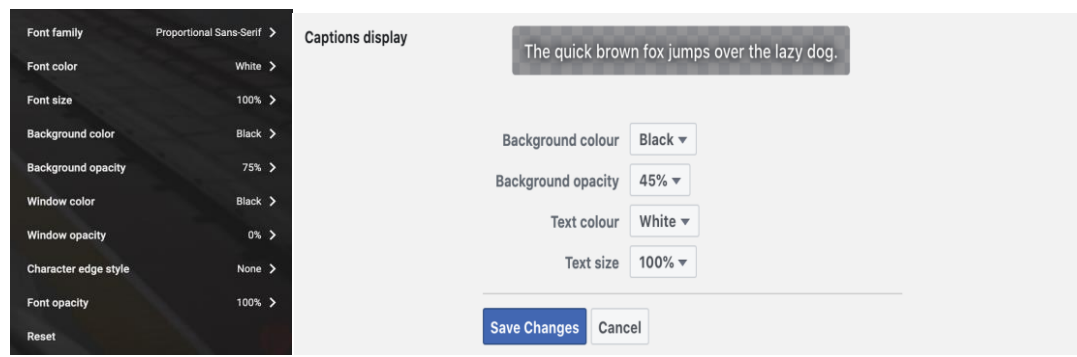

**Figure 3: Customizing the subtitle display in YouTube (left), Facebook (right)**

YouTube provides good support for ST, and although there is no option to customize the position of the ST, it is possible to control the font colour and style as well as the background (see Figure 3, left). YouTube provides a service which attempts to automatically use speech recognition to generate ST for content. Although such a service produces nice results, they may not be good enough for the WCAG, which insists on professionally authored accessibility content. In the 360º video mode, the ST are fixed into the user's view (see Figure 4, left).

---

[2] www.omnivirt.com

[3] www.youtube.com

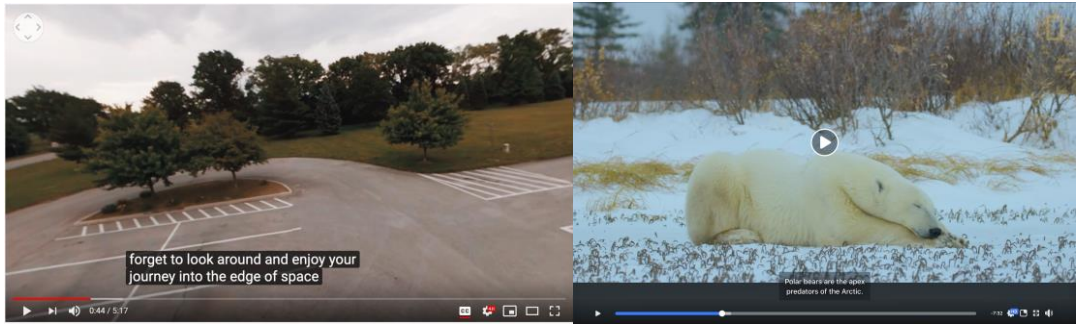

**Figure 4: UIs of the YouTube (Left) and Facebook (Right) 360º players**

Natively, YouTube has no support for AD or SL. It is simply assumed that a user would provide a specific version of the video with a burned in signer video or a specific AD soundtrack. There are however additional projects, such as YouDescribe [17], which allow for YouTube videos to be played through a third party website, whose users collectively provide their own AD for the videos. In this case, the YouTube video is paused at any point that a contributor has specified, and then a text-to-speech engine is used to read the description before resuming the video.

**Facebook**[4] is a large social media platform with over 2 billion active users, and is primarily focused on sharing personal videos and photos. Facebook additionally integrates a 360º player, which does provide limited support for ST, by either uploading an SRT file or using their online tool (which includes an auto generate function based on voice recognition) (see Figure 4, right). There is however no support for AD and SL. Facebook has also adopted a radar symbol to give the user an indication of their orientation within the video, and to re-orientate it towards the centre view. It is possible to set some basic style preferences such as font colour and size (see Figure 3, right). Facebook also takes advantage of users having to be signed in and tracked, which enables the platform to remember the users preferences and also allow the user to specify that ST should be displayed by default.

**THEOplayer**[5] is another 360º player, which consists of a growing portfolio of feature-rich Software Development Kits (SDKs) with wide video ecosystem pre-integration. It supports basic support for ST, including WebVTT and TTML file formats, and the presentation mode consists of fixing the ST in the user's view (see Figure 5, left). The UI shares elements with all the other players, such as the controls at the bottom of the screen, with the play / pause button on the left and the display ST button on the right. It is however only displayed if ST are available, which can leave users searching for this control when it is not displayed. The user has a similar control of the subtitle rendering to JWPlayer and YouTube, and has the ability to change the rendering style but not position.

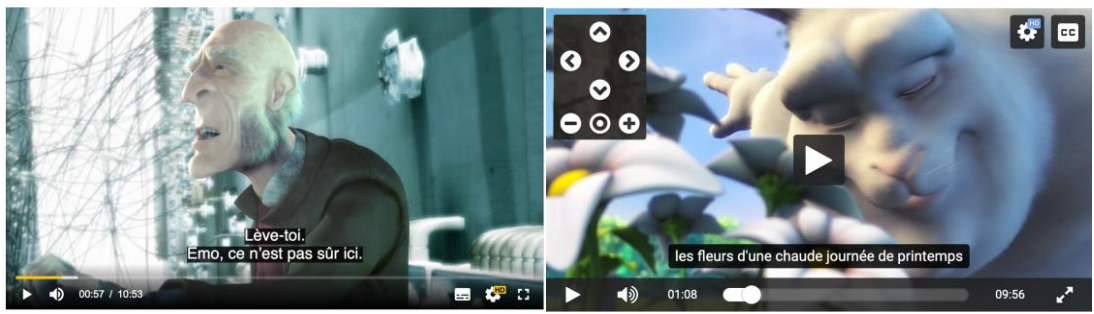

---

[4] www.facebook.com

[5] www.theoplayer.com

**Figure 5: THEOplayer (left), Radiant Player (right)**

Although not directly supporting AD, THEOplayer supports multiple audio and language tracks for one single video – both for live and on demand streaming. The support for multiple audio streams can be then used to provide multiple language tracks for a specific video clip, but also used to provide specific AD tracks. However, switching between audio tracks is needed, which is completely opposed to having a specific AD track in addition to the main audio.

**Radiant Media player**[6] is a commercial video player which is extensive in its implementation. It stands out from the other players by having an enhanced UI (see Figure 5, right), specifically designed for 360º video, which allows the video to be reoriented and zoomed through the interface. It provides a mature support for ST (WebVTT, WebVTT, TTML), but the standard implementation contains no method for the user to customise the display of the ST. However these features can be extended by the video host through the player setup.

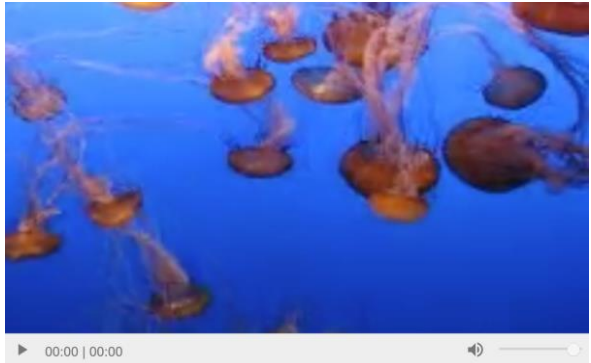

**Figure 6: GoogleVR**

**GoogleVR**[7] is in fact not a 360º player, but an API for developers to create 360º video VR experiences, by Google. One of the main features of this API is the ability to use it in order to directly embed 360º videos onto a web page, which is done via a JavaScript application that creates and controls the contents of an HTML *iframe* element, or by explicitly declaring the *iframe* itself. Although the API supports an extensive selection of file formats, there is no consideration for ST, AD or SL (see Figure 6). In addition, it by default only provides a minimal UI for playing video. However, the general design behind it can be extended to create more advanced UIs, using Javascript and HTML.

## 2.2 Executable 360º video Players

In this subsection, the key 360º video players developed to run as native applications for specific platforms are reviewed with regard to their accessibility features, and being categorized into three main groups based on their primary target platforms: desktop computers, HMDs, and players provided as part of an API.

### 2.2.1. Desktop Computer

360º video players like 5kPlayer[8], VLC [9] and GOM Player[10] (see Figure 7, 8) are designed for running on a desktop computer. By using them, the 360 º video can be viewed in either a window or a full screen, and all interactions with the environment are done with the mouse and keyboard, such as clicking and dragging the video to move the users point of view. These players generally come with support for a wide range of file formats (such as .MPG, .MP4, .WEBM, .AVI, .MOV, .OGG) and share a relatively common UI. Commonly, all of these players come from extensions to existing 2D video players for adding the ability to render 360 º videos, and therefore simply take advantage of the features already provided.

---

[6] www.radiantmediaplayer.com

[7] developers.google.com/vr/discover/360-degree-media

[8] www.5kplayer.com

[9] www.videolan.org

[10] www.gomlab.com/gomplayer-media-player

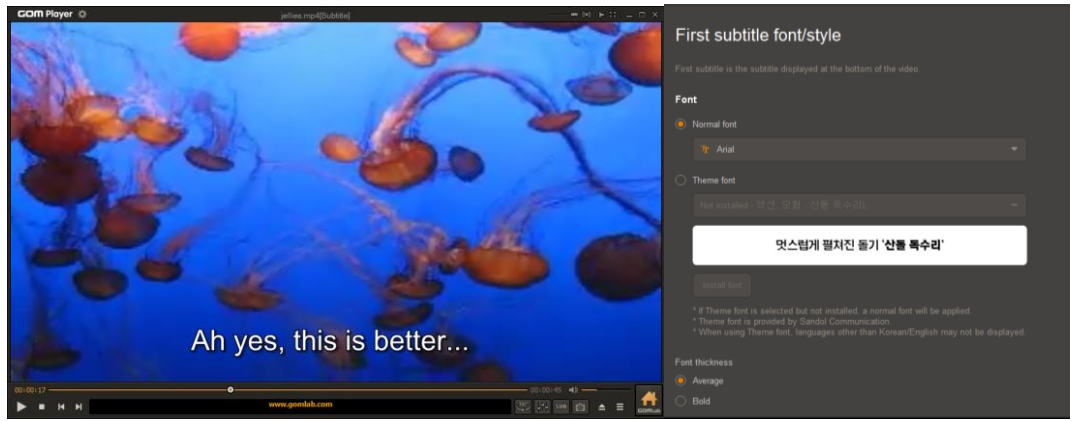

**Figure 7: GOM PLayer: default interface (left), Subtitle Options (right)**

Many of these players are free to use, although **GOM Player** generates revenue by the placement of adverts into the UI, which can be removed by the purchase of a commercial version. **VLC** is the only player that is fully open source, with the software being developed as a community project[11]. These players typically can be run on a wide range of platforms. For instance, **5kPlayer** and **VLC** run on Windows, Mac and Linux. **GOM Player,** however, is only available for Windows. Interestingly, **VLC** has an additional product, '*VLC_VR*', which provides similar functionalities for Android, IOS and Xbox.

All of these players provide some support for ST. Although only VLC allows for the ST position to be moved, they do all allow for the font and style to be changed. **GOM Player** is unique by offering the ability to load two separate subtitle files, with one displayed at the top and the other at the bottom of the screen. However, this support is directly provided by their "predecessors" 2D video players. For example, all of the players allow loading a ST file and rendering it as a 2D overlay onto the video window. As there is no standard ST file specifically designed for 360º videos, there is no special information about where the subtitle relates within the omnidirectional scene. VLC and GOM Player offer some customisation options for the presentation of the subtitles, such as style and position.

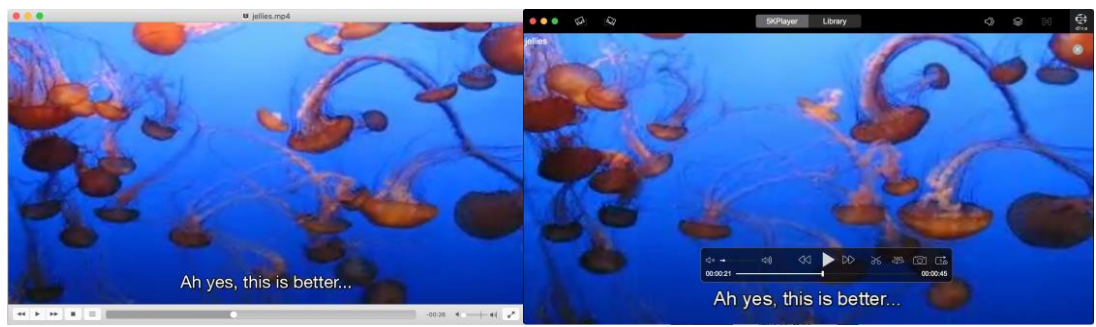

**Figure 8: VLC (left), 5k Player (right)**

These players additionally provide support for selecting alternative audio tracks. These tracks can be used for alternative languages, but could also be used for the AD service. However, there is no mechanism for seamlessly mixing the AD over the existing audio track in the player.

None of the players provide a mechanism for adding SL or overlaying an additional video stream as e.g. in Picture-in-Picture mode, which could be used for the SL service. Therefore, the only way to add a signer would be to burn the additional SL video into the main video, and provide them in a single stream.

---

[11] https://github.com/videolan/vlc

However, it then causes the position of the signer to be fixed and gives no opportunity for the signer position to be fixed in the users view, or to customize its presentation, based on needs and/or preferences.

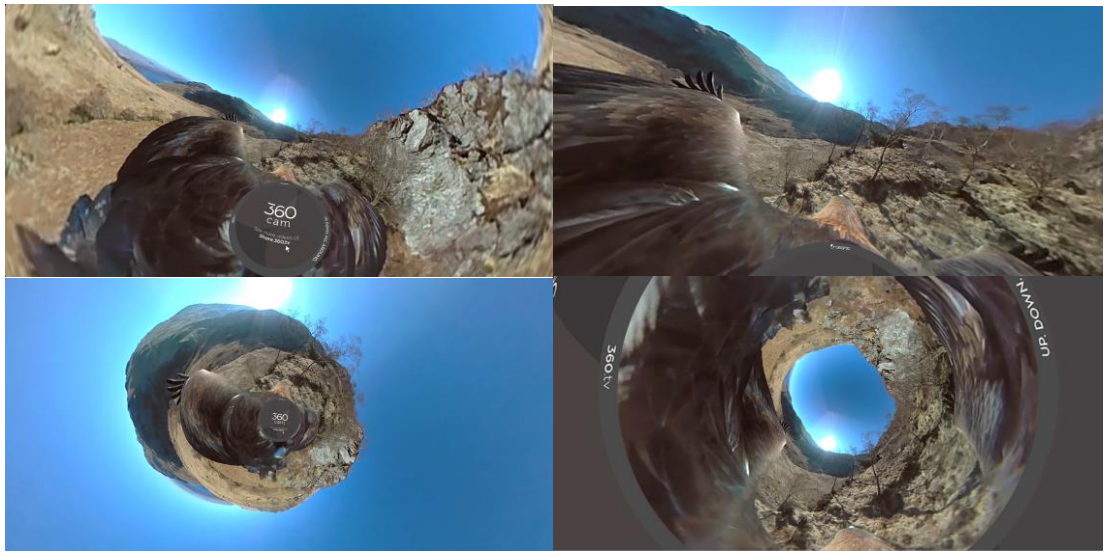

**Figure 9: VLC Rendering modes: Default (top left), Zoomed (top right), Little Planet (bottom left) and Reverse Little planet (bottom right)**

**VLC** is by far the most advanced player in this category. It provides a number of additional rendering modes for 360º videos, such as 'zoomed', 'little planet' and 'reverse little planet' (see Figure 9). It is possible to smoothly scale between these modes, giving partially sighted users further control over how much of the scene they have in view. Although it is primarily a desktop video player, it does also support rendering on connected HMD displays.

Finally, there are other existing players specifically designed for stitching and reviewing footage taken from video cameras, as it is the case of **GoPro VR Player**[12] a Windows application for GoPro cameras. However, GoPro VR Player does not provide any accessibility functionality.

### 2.2.2. HMD

There are two types of HMD's available for consuming immersive VR content. The first type of headsets is 'tethered' to a powerful desktop computer host, where the graphics are rendered and simply displayed on the HMD. This includes HMD's like Oculus Rift, HTC Vive/Pro and PlayStation VR. The second type of headsets is standalone and has its own processing capabilities, like Oculus Go, Oculus Quest and Samsung Gear VR (embedded with smartphones). It is worth noting that each of the standalone HMD's have a web browser provided by their Operating System (OS). This enables the usage of any of the web based 360º players previously reviewed. This is a strong advantage for the web based players [12], as previously highlighted.

**Deo VR**[13] and **VR Player**[14] are two 360º video players designed to run all of the commonly available HMDs (including implementations for HTC, Oculus, Android, IOS and Windows). This is achieved by having native builds for the stand-alone HMD's and for desktop computers acting as hosts. They are both free to use and support basic support for ST and multiple audio streams. The developers for **Deo VR** are aware that its subtitle support is very limited, and it is made clear throughout their support forums that early on, due to limited resources, subtitles were not a priority to them. **VR Player** sets itself apart from other players by integrating successfully with the controllers provided with many of the HMD's. These controllers can be moved freely in space, allowing the user to use gesture control to operate the player. The interface is also extended through both simple voice control and integration with Bluetooth controllers.

---

[12] gopro.com/pt/br/news/gopro-vr-player-2-now-available

[13] deovr.com

[14] vrplayer.com

It is important to note that the existing players in this category, such as VR Player, were in general initially designed for projecting 2D video into the immersive space, by providing a virtual cinema with the video playing on a virtual screen. This is mainly due to HMD's being available sooner than 360º cameras and there being more 2D content available. Other players that operate in this maners include **RiftMax VR**[15] and **Simple VR**[16], which both run under both Windows and Mac hosts with implementations designed for the HTC Vive/Vive Pro. **RiftMax VR** is free and simulates a giant multiplayer Imax-style theater that is able to play 2D and 360º videos. Being a multi-user environment, this means that users can interact with each other and have a shared experience while watching the video. **Simple VR** is a commercial product, but focuses heavily on having an easy to use interface and also allows the use of hand gestures. All of these players are however limited in what they provide for accessibility, and none go beyond the traditional ST rendering.

**Skybox VR**[17] provides a player which runs on standalone headsets. One of the issues faced with the standalone HMD's is that they are fairly limited for storage. **Skybox VR** gets around this by providing a Windows or Mac file server, which streams 360º video files to the headset. It supports all of the common stand-alone HMD's such as Oculus Go, Oculus Quest and Samsung Gear VR. Skybox VR supports traditional ST fixed in the user's view. Although there is no opportunity for customisation, **Skybox VR** provides basic controls for resynchronising the subtitle timing, by delaying or advancing the timings, improving the temporal alignment if needed.

**2.2.3. Development API's**

Several API's designed specifically for building 360º players exist. Commercial groups, such as **Bitmovin**[18], provide complete pipeline solutions for content producers to publish their 360º content in the cloud, by including encoding, analytics and playout solutions. They provide an API for Web, Android and IOS development (see, Figure 10) Although they provide a feature rich API, there is no currently no direct support for accessible services, although implementation seems feasible.

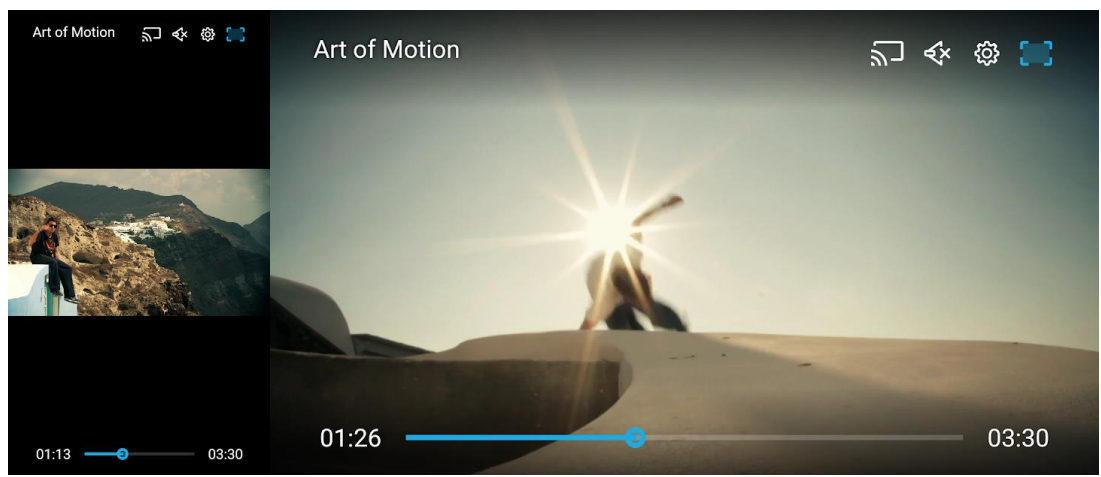

**Figure 10: Bitmovin Player on Android: portrait orientation (left), landscape orientation (right)**

There are other API's, such as **Exoplayer**[19], which provides a direct alternative to Android's MediaPlayer API (see, Figure 11). However, there is no direct consideration for accessibility. Other notable development tools include **Marzipano**[20] which although not being a 360º video player, allows users to create immersive tours from 360º photos, and thus it is worth to be mentioned in this study. However, once again there is no consideration for accessibility, and certainly no opportunity to integrate AD.

---

[15] riftmax.com

[16] simplevr.pro

[17] skybox.xyz

[18] bitmovin.com

[19] exoplayer.dev

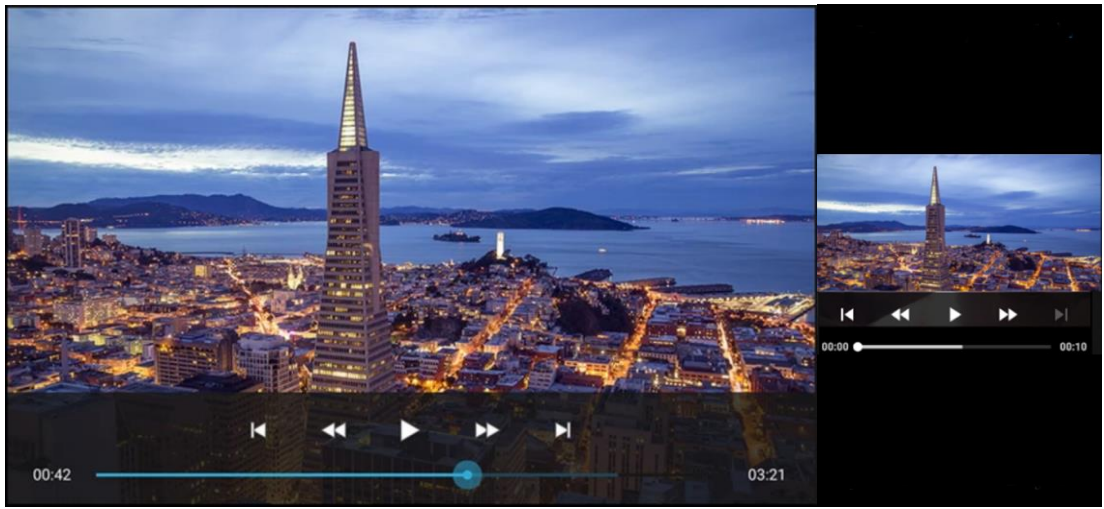

**Figure 11: Exoplayer on Android: portrait orientation(left), landscape orientation (right)**

## 2.3 Ad-hoc players by content providers

In the race to provide innovative and engaging 360º experiences to their audiences, content providers started developing ad-hoc players to offer their produced 360º content. Relevant examples of bespoke web-based player built by content providers are the ones by the BBC[21] (see Figure 12, left), NYT[22] (see Figure 12, right), and RTVE[23] (see Figure 13), developed to meet their specific requirements whilst still trying themselves to understand how 360º video production could be accomplished and massively offered.

Although these solutions do not provide support for loading subtitles, the BBC, NYT and RTVE partially addressed the requirement by adding creative captions burned into the video (see Figures 12 and 13). These captions are not verbatim, and give the user no control on how they are displayed. This does however also mean that the broadcaster can maintain control on the creative look of the ST and guarantee that they will be rendered as designed. It also allows for the designer to provide captions that are fully integrated with the scene in terms of position and style. However, by using this approach, if the caption is burned into the video, it could appear behind the viewer and be missed. Therefore, different broadcasters have addressed the need for navigation in different ways. For example, the BBC took an approach which replicates the ST at 120º intervals around the user, to ensure it is always visible [18], and the NYT adopted a radar in order to give the user spacial awareness as to where they are looking relative to the video. RTVE additionally developed a VR app for the consumption of their produced 360º videos (see Figure 13, right).

More recently, with the widespread adoption of 360º video services, content providers, including the mentioned ones, have commonly moved towards using major full-fledged platforms, like YouTube, for hosting and distributing their 360º videos to their audiences. An example is provided in Figure 14 for a branded version of the Youtube player for RTVE.

---

[20] marzipano.net

[21] https://www.bbc.com/reel/playlist/reel-world-360-videos-from-the-bbc

[22] https://www.nytimes.com/2016/11/01/nytnow/the-daily-360-videos.html

[23] https://www.rtve.es/lab/realidad-virtual/

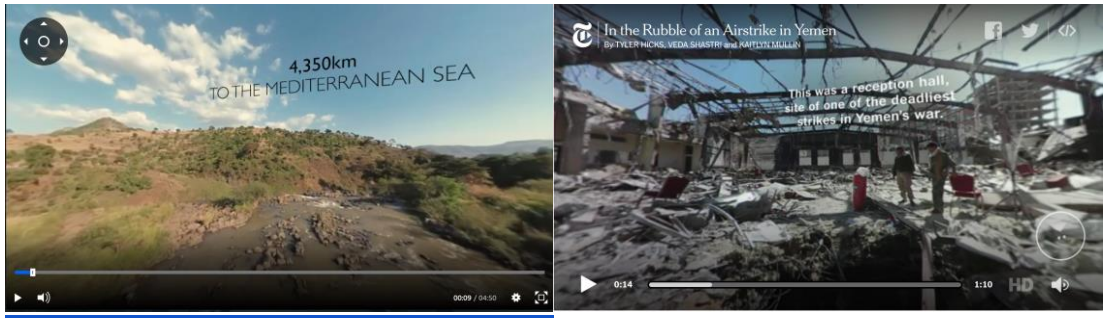

**Figure 12: BBC player (left), NYT player (right)**

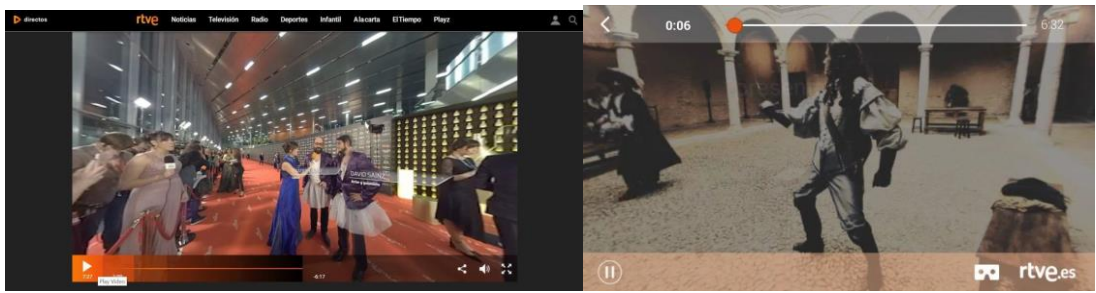

**Figure 13: RTVE player: web based player with creative textuals burned in (left), VR app (right)**

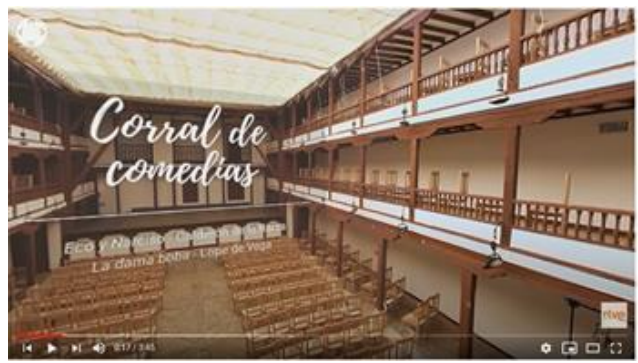

**Figure 14: RTVE using a branded YouTube player, with textual burned in (attached to the 360º scene)**

# 3 Access to Access Services

In order for an access service to be effective and useful, it needs to be easy to access, and interact with. This is particularly relevant to users with visual impairments. From this study, it becomes clear that there is no standard and unified approach to designing the 360º player UI, including the controls for accessing the access services and for setting the available features. In addition, it is often not clear if the access services are available at all. For example, in all reviewed players, if subtitles are not provided/supported, the activation button simply does not display, causing the user to search around for the options. These findings are inline with the ones in [19], mostly focused on reviewing traditional 2D players.

Generally, as shown in Figure 15, the controls are positioned at the bottom left of the video window. However, there are no standards to define this, so developers typically position the controls anywhere they choose. For example, the menu and service controls are positioned at the top right of the screen in the Radiant Media Player (Figure 15, left).

The positioning of the controls at the bottom or top is generally a good practice in 2D players to provide a cleaner interface for video viewing, minimizing blocking issues caused by the UI. These control positions are typically mapped into the VR mode of these platers, replicating the same locations. However, this is not an appropriate approach, especially when using HMDs. This is due to the fact that the HMD's typically have a limited Field-of-View (FoV), ranging from 90º-110º, and placing visual elements around the edges of the FoV leads to uncomfortable viewing experiences [4].

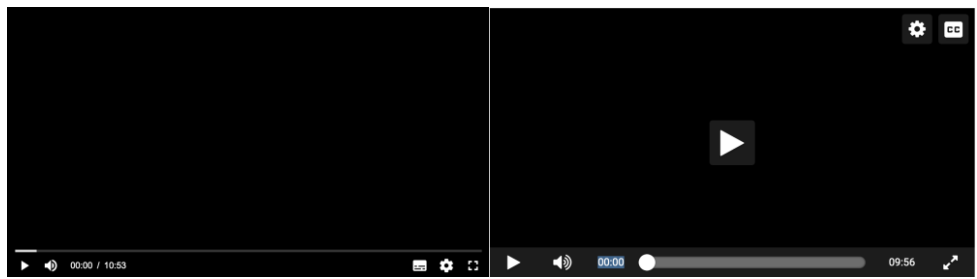

**Figure 15: Typically the controls to open the access services are located on the bottom right such as in THEOplayer (left), with exceptions such as Radiant Player which positions the controls top right (right)**

With regard to UI elements, there is no standardized solution, although similar approaches have been followed on web based players. As shown in Figure 16, different symbols have been chosen for the ST service, which is made even more confusing by the common choice of a 'CC', taken from the US 'Closed Caption' description. The use of the service acronym is sometimes also used for this, which besides varies depending on the active UI language. In addition, while this is already an issue for ST, it becomes more evident and diverse for the less developed access services, like AD and SL, which although being very scarcely supported in 360º players, their support is a bit more frequent in traditional 2D players [19].

| JWPlayer | YouTube | Facebook | TheoPlayer | Radiant Media Player |
|---|---|---|---|---|
| CC ⚙ | CC ⚙ HD | CC ⚙ | ▭ ⚙ | ⚙ CC |

**Figure 16: Examples of the button designed for opening subtitles.**

The executable 360º players make it even more difficult to access the ST service. For example, in a desktop environment, both **5K Player** and **VLC** require the user to select subtitles through the menu as shown in Figure 17. Although **VLC** does offer a keyboard shortcut, this key varies between OS, and can be hard to identify if the correct subtitle track has been selected.

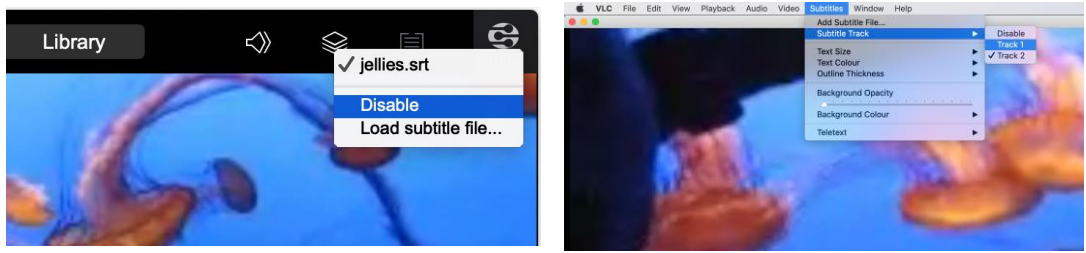

**Figure 17: Selecting subtitles on a desktop 5K Player (left) and VLC (right)**

Most of the executable players also replicate the desktop UI in the HMD mode. As shown in Figure 18, the interface for selecting ST on the **Skybox VR** player is very similar to those found in the other desktop applications we surveyed. However, it is significantly more difficult to control a menu within an immersive HMD environment, making it even more challenging to find the access services controls. Other players, like **Deo VR,** offer no control for ST. Using **Deo VR** player, the ST are loaded and displayed by default, if available. Although this offers a direct solution to users requiring ST, it could be annoying to users who prefer to not use them, and no presentation control is provided.

Another relevant aspect in this context is to minimize the number of clicks/interactions required to (de-)activate the access services and to set the available personalisation options. Whilst typically 2D players provide quite efficient solutions for these purposes [19], these issues do not yet apply to 360º players, due to their scarce support for access services presentation and personalisation. However, they need to be taken into account when addressing these gaps/requirements.

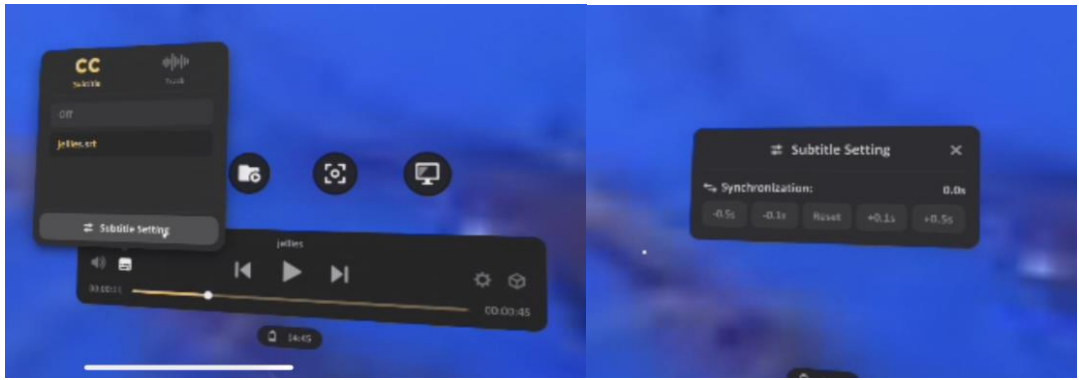

**Figure 18: Selecting and controlling subtitles on SkyboxVR in HMD mode**

Due to the diversity in terms of UI elements and icons for the access services [19], DR Design has proposed four universal icons (Figure 19) to identify access services, to be adopted as a standardized solution for player UIs [20]. By standardizing, the icons would not only become easily recognisable, but also independent of any specific language, preventing non-native language speakers from using the services. The Danish Broadcast Corporation (Figure 20) has already started adopting the icons for their content, and the European Broadcasting Union (EBU) through its Access Services Group of Experts is supporting this initiative for standardized universal icons. Since the icons are text-based, they can be typed on a keyboard, included in descriptive metadata and read out aloud (e.g., using a screen-reader). The ultimate goal is to increase simplicity, quality and usability through standardization.

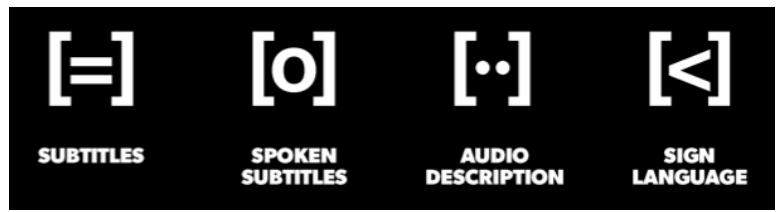

**Figure 19: A standard approach to identifying accessibility services, proposed by DR Design [20]**

Finally, having a traditional graphical menu may not be appropriate at all for visually impaired users. The use of assistive methods in this context can contribute to a better accessibility. Examples are visual feedback when navigating over the menu options, when setting the desired features/options, as well as magnification features (e.g. an enlarged version of the menu, or magnifying visual elements once having the focus). Beyond some basic visual feedback features, the available 360º players do not fully address these important issues.

Due to the possible limitations with regard to visual interactions, the players need to adequately inter-operate with existing screen-readers. This is commonly not an issue for web-based and executable 360º players when running on non VR mode, thanks to the use of standardized metadata that is typically recognised successfully by screen-readers. However, this is an issue in the VR context, and the ad-hoc development of bridges between the screen-reader and the VR engine, or of specific screen-reader like features. Similarly, voice control becomes a very useful interaction modality. Despite initial efforts towards the integration of this feature in existing players (not just at the platform level), like for the **JWPlayer** [16] and **VR Player**, a full integration is not yet commonplace in typical players, unfortunately.

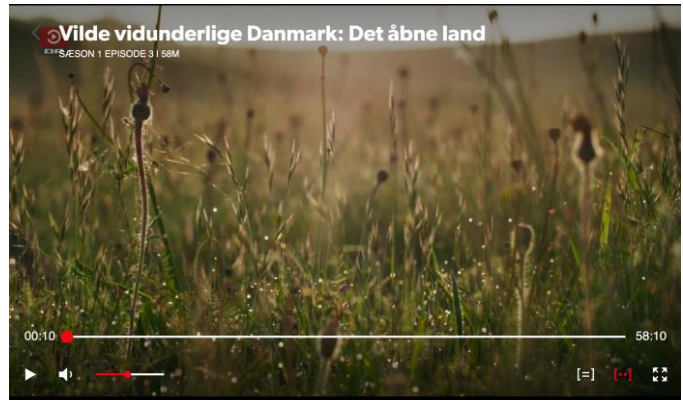

**Figure 20: Standardized symbols used by the Danish Broadcasting Corporation (2D player)**

# 4 ImAc Player

As a response to the gathered insights and identified limitations in the conducted survey and categorization, a cross-disciplinary team has worked together, under the umbrella of the EU H2020 ImAc project[24], to enable an efficient integration of accessibility solutions within immersive 360º media services. In particular, the ImAc project consortium has brought together nine organisations with a rich expertise in different key fields:

- Two Research centers, with experts in the broadcast sector, multimedia systems, and User Experience (UX).
- Two Universities, with experts in interactive technologies, media accessibility, translation and UX.
- Two Companies, with recognised engineers in the media and broadcast sector.
- Two Broadcasters, with a deep knowledge about the audience's needs and about how to deploy real media service.
- An End-User Association in the accessibility field, allowing to gather accurate requirements and to validate them with users.

By combining the expertise and efforts of this team, the consortium has developed a modular end-to-end toolset to allow the integration of immersive and accessibility content in current broadcast-related services [2], including the necessary components from media authoring to media consumption. A key component of the ImAc platform is the open-source web-based 360º player, which enables an interactive and hyper-personalized presentation of access services features for 360º content, combined with a set of assistive technologies. The specific details about the player are provided in [1, 2, 4], but they are also briefly reviewed in this section for the sake of comparison with the surveyed players.

## 4.1 Research Methodology

The integration of accessibility solutions into new technologies from the start contributes to a more effective deployment and exploitation. Based on this premise, the process towards the design and development of the ImAc player has been built on three key pillars: 1) requirements gathering, 2) development and integration; and 3) validation and dissemination. In such a process, a *user-centric methodology* has been followed to accurately gather the accessibility, interaction and personalization requirements. This in turn requires the involvement of end-users, professionals and stakeholders at every stage of the project, through the organization of workshops, focus groups, tests, and the attendance to events, thus closely adopting the "*Design for users with users*" motto.

---

[24] https://www.imac-project.eu/

In particular, the design and development process of the ImAc player have undergo two iterative cycles, including the following key activities:

- **Focus groups** with end-users to derive users' needs and preferences in terms of access service presentation modes, personalisation features, and interaction modalities. Examples are focus groups conducted for ST [21] and AD [22].
- Novel **technological contributions** to implement these required features. Published results in this context include e.g. [1, 4, 23].
- **User testing** to validate and refine the technological contributions with professionally produced content. Published results in this context include e.g. [24, 25].

## 4.2 Technology

In addition to the user-centric methodology, a key premise has been adopted in the development of the player: to guarantee backward-compliance with current formats, technologies, infrastructures and practices in the broadcast/media ecosystem. This will maximise re-usability, interoperability and the changes of successful deployment and exploitation.

In particular, the technology developed to allow the signaling of access services and their appropriate and personalised presentation in 360º environments is based on extending the MPEG DASH[25] and W3C Internet Media Subtitles and Captions (IMSC) subtitles[26] standards. Likewise, the player has been built by making use of HTML5 and Javascript, and by adopting widespread web components and APIs, such as: *dash.js*[27] (the reference player for DASH), *three.js*[28], WebXR[29], and IMSC rendering libraries[30]. In addition, the player includes the necessary technological solutions to enable multi-screen scenarios in a synchronized and interactive manner, in both fully web-based and Hybrid Broadcast Broadband TV (HbbTV) [26] scenarios, as described in [1].

The use of web technologies and components guarantees cross-device, cross-platform, and even cross-browser support, which means that the player can be effectively on traditional consumer devices (e.g. Connected TVs, PCs, laptops, tablets and smartphones) and on VR devices (e.g. HMDs).

A demo video showcasing the ImAc player features can be watched at: https://bit.ly/2Wqd336 Its current version and a wide sample of 360º videos with access services can be accessed via this URL: http://imac.i2cat.net/player/ Finally, the source code can be downloaded from: https://github.com/ua-i2cat/ImAc

## 4.3 ImAc player UI

The ImAc player includes a landing page, i.e. portal, for the selection of the available 360º content, and for initially personalizing the media experience (see Figure 21).

Once a video is selected for playout, the player menu is run. The player menu (see Figure 22, left) can be opened by a single click, looking down for a period, or via voice control. The menu has been designed to be adapted for all range of potential consumption devices (TVs, desktop computers, smartphones, HMD's…), taking special care on the limited FoV in smartphones and HMDs in order to provide a comfortable viewing experience. This issue is one of the key lessons learned during the first round of tests, in which a menu expanding all over the FoV was designed (see Figure 22, right). With regard to the access to the access services, the menu has adopted the introduced universal icons (see Figure 22, left), unlike its first version that added acronyms that depended on the active UI language (see Figure 22, right) [4]. The menu also provides visual feedback when interacting with it (see Figure 24, left) and to indicate the current settings (see e.g. the activated ST service, the menu section with the focus, and the magnification features for the menu in the screenshots in Figure 23, left). In addition, that magnifier control, which is the element at the top left of the menu, allows opening an enlarged version of the menu (Figure 24, left), which is more suited for users with sight loss, and also when using small screens, like in smartphones. This enhanced-accessibility variant of the menu has also significantly evolved since its first version presented in [4] (Figure 24, right). For blind users, the player includes a voice control feature, by having developed a gateway that communicates with Amazon Echo (i.e. Alexa).

---

[25] https://mpeg.chiariglione.org/standards/mpeg-dash/

[26] https://www.w3.org/TR/ttml-imsc1.0.1/

[27] https://github.com/Dash-Industry-Forum/dash.js/wiki

[28] https://threejs.org/

[29] https://www.w3.org/TR/webxr/

[30] https://github.com/IRT-Open-Source

Finally, as the Omnivirt and Facebook players which include a radar to indicate where the centre of the scene is, the ImAc player also includes guiding methods to assist the users in not losing track of the main action, additionally indicating where the active speaker is in the 360º space. This is provided by means of a radar or arrows, as preferred by the users (see Figure 23).

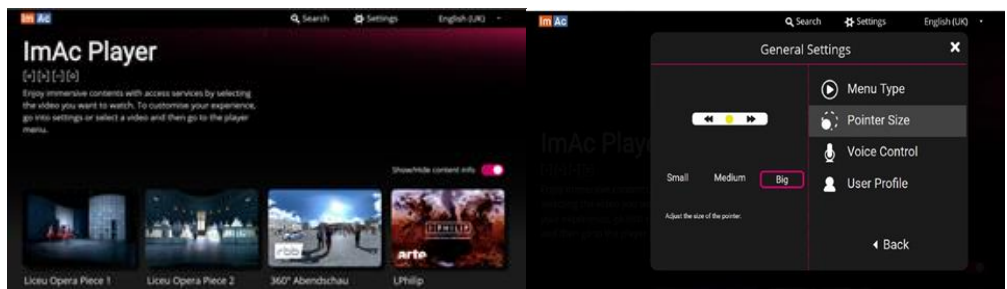

**Figure 21: The ImAc portal UI**

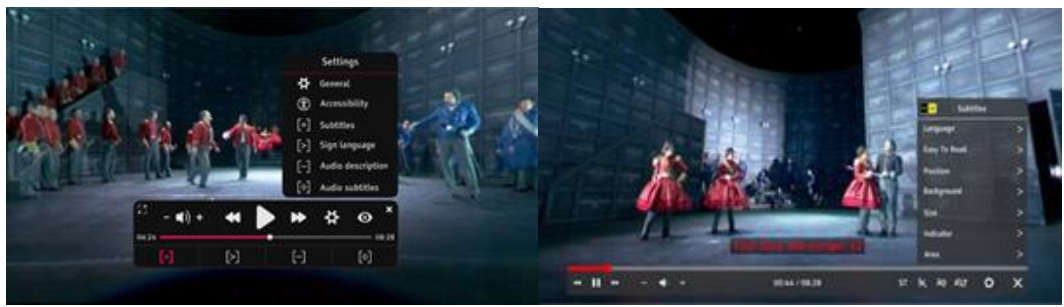

**Figure 22: The ImAc player UI (latest version on the left; first version on the right)**

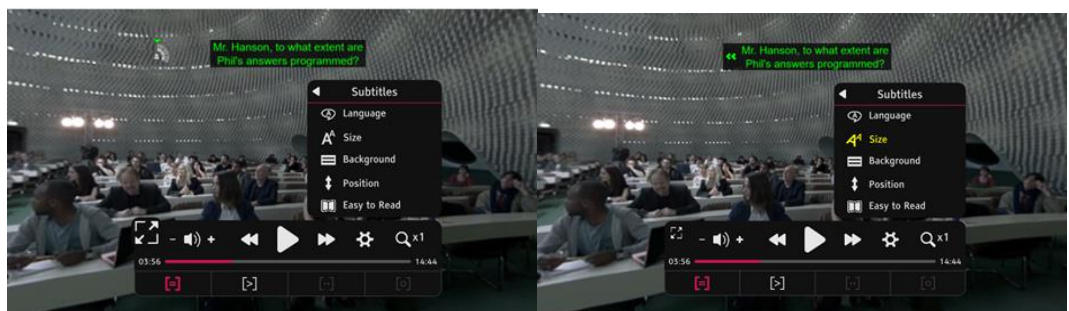

**Figure 23: Visual Feedback when interacting with ImAc player UI**

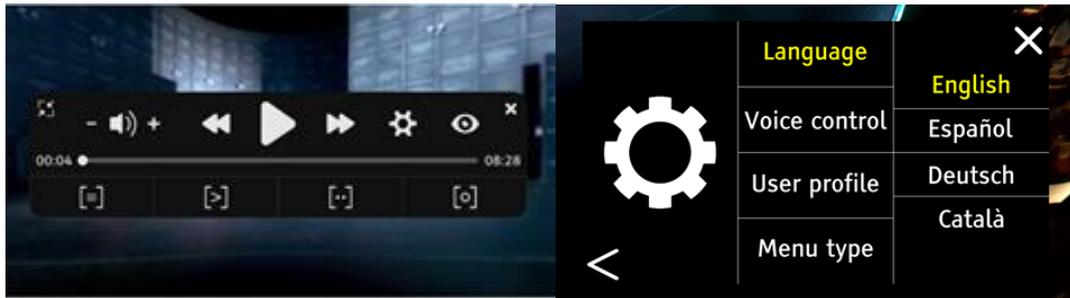
**Figure 24: Enlarged version of the ImAc player UI (latest version on the left; first version on the right)**

## 4.4 Presentation of Access Services in the ImAc player

**Subtitles (ST)**

Unlike the existing 360º players that mainly present ST fixed in the user view, the ImAc player allows different presentation modes for ST:

- Fixed in the user view (aka always-visible ST), as most of the existing players;
- Fixed to scene, by replicating the ST at 120º intervals around the user, as done by the BBC [18].
- Fixed to speaker [23], which is similar to the above mode, but by rendering the ST close to the associated speaker, and additionally using alway-visible guiding methods (e.g. arrows) to indicate where the speaker is, if he/she is outside of the user's FoV. As long as the speaker is within the user's FoV, the visual indicator is automatically hidden. This presentation mode is outlined in Figure 25.

Conducted tests in [24, 25] have shown that alway-visible subtitles are clearly preferred, mainly because they were easier to find and to read, less distracting, and users perceived a higher freedom to explore the 360º environment without missing the subtitles.

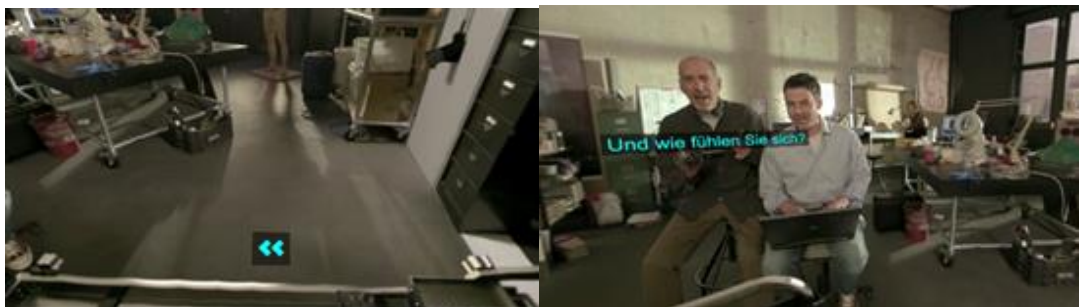
**Figure 25: Subtitles attached to the speaker with always-visible visual indicators**

With regards to the indicators, it is worth mentioning that: 1) the arrows are only shown if the speaker is outside the FoV and are, and are automatically hidden when the speaker is again visible; 2) the radar indicates the current user's FoV and the relative position of the speaker, by using a mark of the same color as the subtitles for a better identification.

Unlike the reviewed 360º player which provide few personalisation options, except Youtube, the ImAc player allows a personalised presentation of ST in terms of: size (three size levels); background (outlined text or a semi-transparent background box); position (top and bottom); and language.

Finally, the presentation of Easy-to-Read subtitles is supported. Results from conducted tests have preliminarily proven that Easy-to-Read subtitles are preferred over traditional subtitles by elderly participants when watching 360º clips from an opera performance [27].

**Sign Language (SL)**

The ImAc player supports the presentation of SL, but not as a burned in video, as in the scarce solutions supporting SL, but as an independent DASH stream signalised as part of the main 360º video service. On the one hand, this allows presenting the SL video fixed in the user view, and not fixed to the scene, thus being always visible regardless of where the user is looking at. On the other hand, this allows dynamic personalisation of the SL service, in terms of activation/deactivation, and of language, size (three levels), and position settings (left, right). Two visual indicators (arrows and radar) can be also enabled, as for subtitles (see Figure 26). In order to provide a better identification of the target speaker, his/her name (or even a descriptive info text) can also be added below the video window (see Figure 26).

Two further innovation features of the ImAc player can be also highlighted:

- It enables the simultaneous presentation of ST and SL. In such a case, if the subtitles are moved at the top via the associated option of the player menu, then the sign language videos will be also moved at the top for a better visual alignment between both access services. Likewise, ST is considered as the master service for indicators.
- It allows dynamically showing/hiding the video window based on the signer's activity, based on metadata added at the production side.

Finally, all visual elements on screen, including ST, SL and indicators, can be dynamically placed at the preferred position, thanks to a developed drag & drop feature (see Figure 26, left, where the radar is being moved, and this is indicated by a yellow outline.

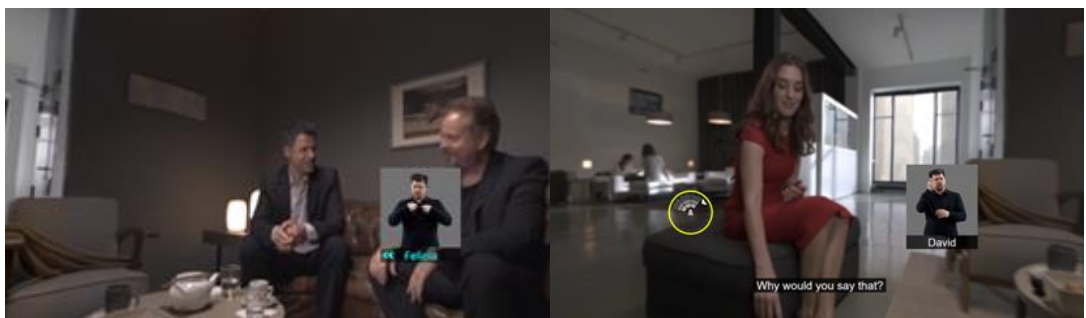

**Figure 26: Presentation of SL with indicators, and simultaneously with ST (right)**

**Audio Description (AD)**

The ImAc player also provides does not just provide support for AD as independent streams, but leverages the availability of spatial audio technology (Ambisonics) to provide different forms of presentation modes and narratives in a personalized manner (if available), like:

- Classic Mode: no audio positioning.
- Static Mode: audio from a fixed point in the scene (e.g. like a friend whispering in your ear).
- Dynamic Mode: audio coming from the direction of the action.

Likewise, the fact that multiple independent streams allows for: 1) adding different scripting and narrative modes; 2) dynamic personalization in terms of language, presentation mode and volume levels.

The same features are supported for audio subtitles (AST), which is a much less developed access service, but that can provide very relevant features [28].

# 5 Taxonomy and Conclusions

This study has reviewed the key existing 360º with regard to the relevant accessibility related aspects. This overall review is categorized in Table I to show an overall picture of to what extent the identified accessibility guidelines are met, and what the existing gaps are.

**Table I: Summary of directly supported accessibility in key 360º video players**

| | | Subtitle | Audio Description | Sign Language | Accessible UI | Voice Control | Multi screen | VR Mode |
|---|---|---|---|---|---|---|---|---|
| **Web based** | JWPlayer | ✓ | ✓ | | | ✓ | | ✓ |
| | Omnivirt | | | | | ✓ | ✓ | ✓ |
| | Youtube | ✓ | ✓ | | | | | ✓ |
| | Facebook | ✓ | | | | | | ✓ |
| | TheoPlayer | ✓ | | | | | | ✓ |
| | Radiant Media Player | ✓ | | | ✓ | | | ✓ |
| | Google VR | ✓ | | | | ✓ | | ✓ |
| | **ImAc Player** | ✓ | ✓ | ✓ | ✓i | ✓ | ✓ | ✓ |
| **Executable** | 5KPlayer | ✓ | | | | ✓ | | |
| | VLC 360 | ✓ | | | | | ✓ | |
| | GOM Player | ✓ | | | | ✓ | | |
| | Skybox VR | ✓ | | | | | | ✓ |
| | Simple VR | ✓ | | | | | | ✓ |
| | Deo VR | ✓ | | | | | | ✓ |
| | Bitmovin | ✓ | | | | | | |
| | Exoplayer | ✓ | | | | | | |
| | RiftMax VR | ✓ | | | | | | ✓ |
| | VR Player | ✓ | | | | ✓ | | ✓ |
| | LiveViewRift VR | | | | | | | ✓ |
| | GoPro VR Player | | | | | | | |
| | Marzipano | | | | | | | |

[i] Apart from an adaptive and responsive interface, the ImAc player UI provides multi-language support and assistive methods, like visual feedback and magnification features while interacting with the menu

In general, the existing 360º players provide very little support for accessibility. The majority of players do support ST, but only as a traditional television rendering into the 360º projection. This means that they have some mechanism for displaying text which can be turned on and off, however there is no consideration for the space where it is to be rendered or its location within the scene. This is generally because they follow the principles used within traditional television broadcast, providing two lines of text, ~30 characters wide. There is also no mechanism within standard subtitle file formats to store position information, which would allow implementing a logic to indicate where the active speaker is in the 360º space at any moment, relative to the user's FoV.

There is absolutely no support for SL as a separate stream within any of the existing players. It has been generally accepted that if a signer is required, it can be burned into the video. However this gives no scope for customisation, such as allowing the user to turn the feature on or off, customizing position and size or keeping the signer within the users viewpoint. In 360º video this can be a serious limitation, as the producer of the video has to choose where to put the signer and therefore forcing the user to follow a specific view. In case of actions happening around the 360º scene, it could also mean that the SL had to be re-located or that some relevant scenes could be blocked by the SL video.

Some players provide support for AD, but only as an alternative audio track, not an additional sound track which can be balanced against the main audio. This causes a major disadvantage for those with hearing impairments, as being able to control the AD level against the background can massively improve their experience. The only 360º players that natively support AD as an additional audio stream are JW Player and Youtube. However, as far as authors know, they do not include specific personalisation options for AD, beyond those common for audio. In this context, no 360º players providing support for AST exist.

Although some of the players, such as JWPlayer, provide an API for developers to implement their own UI, there is very little consideration for the default UIs to be adaptive to meet the specific needs for: an omnidirectional environment projected onto a sphere; the limited FoV in HMD; and accessibility. This includes the ability to extend the media experience onto multiple screens. This is only provided by Omnivirt, thanks to community developed extensions.

Voice control is becoming increasingly popular, mainly due to the expanding market for voice activated digital assistants (such as Amazon Echo, Google Home and Apple Siri), and the advantages to partially sighted users are clear as it negated the need to directly see or operate the UI. Currently, there are available mechanisms to connect these digital assistants to most platforms, connecting at the system level. This can allow some control, such as starting the video player. However, the lack of standards for connecting the voice control to the UI prevents from actually providing an effective control over the actual player.

Finally, it was found that generally the players are free to use. Although the development costs are high, revenue is normally generated through both advertising and the sales and licensing of the authoring tools.

In general terms, the surveyed 360º players hardly meet the necessary accessibility requirements, and the provided ones seem to be inherited from the traditional 2D world, instead of addressing the specificities of 360º environments. In terms of the WCAG guidelines, the existing web-based 360º players still present important limitations.

All these limitations have been overcome by the ImAc player, which has successfully addressed the recommended accessibility requirements and beyond.

# 6 Future Work

The contributions and insights from this study can be used as a catalyst to improve the existing solutions, and to adopt the best suited ones by the interested agents (end-users, developers, content providers…). They can be also used to support standardization activities and future research efforts, thanks to the provided overall view of to what extent this research space is encompassed. Likewise, the development of the ImAc player is not meant to be completed. Its developers will continue monitoring the research topic and new opportunities to further refine and extend it. This can also be done by third-party agents, as the player is open source and released on Github.

Finally, a similar research study should be also conducted for 3D VR environments, with 6 Degrees of Freedom (6DoF), where additional challenges need to be addressed. The work in [29] is a starting point in that direction.